\documentclass{azjmPREarxiv}

\def\cl{{\cal C}\!\ell}
\def\R{{\mathbb R}}
\def\C{{\mathbb C}}
\def\F{{\mathbb F}}
\def\T{{\rm T}}
\def\Tr{{\rm Tr}}
\def\Spin{{\rm Spin}}
\def\SO{{\rm SO}}
\def\Or{{\rm O}}
\def\Cen{{\rm Cen}}
\def\Mat{{\rm Mat}}

\begin{document}

\title{Local generalization of Pauli's theorem}

\author[ N. G. Marchuk, D. S. Shirokov]{ N. G. Marchuk, D. S. Shirokov\corrauth}

\begin{abstract}
Generalized Pauli's theorem, proved by D. S. Shirokov for two sets of anticommuting elements of a real or complexified Clifford algebra of dimension $2^n$, is extended to the case, when both sets of elements depend smoothly on points of Euclidean space of dimension $r$. We prove that in the case of even $n$ there exists a smooth function such that two sets of Clifford algebra elements are connected by a similarity transformation. All cases of connection between two sets are considered in the case of odd $n$. Using the equation for the spin connection of general form, it is shown that the problem of the local Pauli's theorem is equivalent to the problem of existence of a solution of some special system of partial differential equations. The special cases $n=2$, $r\geq 1$ and $n\geq 2$, $r=1$ with more simpler solution of the problem are considered in detail.
\end{abstract}



\maketitle

\section{Introduction}

In the paper \cite{Shirokov}, one of the authors presented statements describing connection between two sets $g^a$, $h^a$, $a=1, \ldots, n$ of elements of a real or complexified Clifford algebra $\cl^\F(p,q)$, $p+q=n$, that satisfy the relations
$$g^a g^b+g^b g^a=2\eta^{ab}e,\qquad h^a h^b+h^b h^a=2\eta^{ab}e,\qquad a, b=1, \ldots, n,$$
where $\eta^{ab}$ is the diagonal matrix with $p$ pieces of $1$ and $q$ pieces of $-1$ on the diagonal.
These statements generalize Pauli's theorem proven for the case $n=4$ \cite{Pauli}. We call these statements \emph{algebraic generalized Pauli's theorem} and actively use them in the study of spin groups \cite{spin} and $n$-dimensional spinors \cite{tmf}.

In the present paper, we generalize these statements to the case when both sets of elements depend smoothly on the points of Euclidean space. We call these statements \emph{local generalized Pauli's theorem}.

First we show that connection of two sets is realized locally, in the neighborhood of the point of Euclidean space. Secondly we generalize the statement to the case of the entire Euclidean space under certain assumptions. We also call the obtained statement \emph{local}, as is customary in differential geometry (statement is called \emph{global} if it holds for the entire non-trivial manifold, while Euclidean space is a trivial manifold because it is covered by one chart).

\section{Local generalized Pauli's theorem in the neighborhood of the point of Euclidean space}

Let $V$ be $r$-dimensional Euclidean space with a scalar product $(x,y)$, $\forall x,y\in V$ and the norm
$$
\|x\|=\sqrt{(x,x)},\quad \forall x\in V.
$$

Let $\Omega$ be an open domain in $V$ and let $\varepsilon>0$ be a positive real number, $\varepsilon$-neighborhood of a point $x_0\in V$ is the domain
$$
O_\varepsilon(x_0)=\{x\in V\ :\ \|x-x_0\|<\varepsilon\}.
$$

Let us consider the real Clifford algebra $\cl^\R(p,q):=\cl(p,q)$, $n=p+q$ with the generators $e^a$, $a=1,\ldots,n$, and the basis of $2^n$ elements
\begin{equation}
e, e^a, e^{ab},\ldots, e^{1\ldots n}, \label{eebasis}
\end{equation}
enumerated by the ordered multi-indices of the length between $0$ and $n$. The identity element of $\cl^\R(p,q)$ is denoted by $e$. The generators satisfy the anticommutative relations
$$e^a e^b + e^b e^a =2\eta^{ab}e,\qquad a, b=1, \ldots, n,$$
where $\eta^{ab}$ are elements of the diagonal matrix $\eta$ of order $n$ with $p$ pieces of $1$ and $q$ pieces of $-1$ on the diagonal.

Let us consider the complexified Clifford algebra $\cl^\C(p,q):=\C\otimes\cl^\R(p,q)$ \cite{Lounesto}, \cite{tensor}. These two cases $\cl^\F(p,q)$, $\F=\R, \C$ are important for various problems of mathematical physics, in particular, in the study of the Dirac equation and the Yang-Mills equations \cite{Marchukbook}, \cite{marchuk1}, \cite{ofe}.

The subspace of $\cl^\F(p,q)$ spanned over the basis elements enumerated by the ordered multi-indices of length $k$ is denoted by $\cl^\F_k(p,q)$ and is called {\em the subspace of grade $k$}. We have
$$\cl^\F(p,q)=\bigoplus_{k=0}^n \cl^\F_k(p,q).$$

Note that we have the important special case $r=n$, when the basis of $V$ is the set of generators $e^a$. In this case, $V$ can be considered as pseudo-Euclidean space with two metrics (Euclidean and pseudo-Euclidean). We use Euclidean metric to determine a neighborhood of a point.

Consider a function
$$
f : \Omega \to \cl^\F(p,q)
$$
with values in $\cl^\F(p,q)$. The function $f=f(x)$ can be written in the form
$$
f = u e + u_a e^a + \ldots + u_{1\ldots n} e^{1\ldots n},
$$
where $u=u(x)$, $u_a=u_a(x)$, $\ldots$ are functions $\Omega\to\F$ and the basis elements (\ref{eebasis}) do not depend on $x\in V$.

If (real or complex) functions $u,u_a,\ldots,u_{1\ldots n}$ have continuous derivatives up to order $k$ in $\Omega$, then we say that functions $u,u_a,\ldots,u_{1\ldots n}$ and $f$ belong to the class $C^k(\Omega)$ ($C^0(\Omega)$ is the class of continuous functions in $\Omega$).

\begin{theorem}{\rm (The case of even $n$).}\label{th1}

Let $n$ be an even positive number and $h^a=h^a(x)$, $g^a=g^a(x)$, $a=1,\dots,n$ be functions $\Omega\to\cl^\F(p,q)$ of the class $C^k(\Omega)$ such that
\begin{eqnarray}
&&h^a(x) h^b(x) + h^b(x) h^a(x)=2\eta^{ab}e, \quad a,b=1,\ldots,n,\quad \forall
x\in\Omega,\nonumber\\
&&g^a(x) g^b(x) + g^b(x) g^a(x)=2\eta^{ab}e, \quad a,b=1,\ldots,n,\quad \forall
x\in\Omega.\nonumber
\end{eqnarray}
Then for any point $x_0\in\Omega$ there exist $\varepsilon>0$ and $T=T(x) : O_\varepsilon(x_0)\to\cl^\F(p,q)$ such that
\begin{enumerate}
\item $T(x)$ is a function of the class $C^k(O_\varepsilon(x_0))$;

\item $T(x)$ is an invertible element of $\cl^\F(p,q)$ for any point $x\in O_\varepsilon(x_0)$;

\item $g^a(x)=T^{-1}(x)h^a(x) T(x)$, $a=1,\ldots,n$, $\forall x\in O_\varepsilon(x_0)$;

\item The function $T(x)$ is defined up to multiplication by (real in the case $\F=\R$ and complex in the case $\F=\C$) function of the class $C^k(O_\varepsilon(x_0))$ that is not equal to zero for any point of $O_\varepsilon(x_0)$.
\end{enumerate}
\end{theorem}

\begin{proof} Let us consider the special case of the statement, when the elements $g^a=g^a(x)$ are equal to the generators $e^a$ of $\cl^\F(p,q)$, which do not depend on $x$. First we prove the theorem for this case.

We denote the elements of the basis (\ref{eebasis}) by $e^A$, where $A$ are ordered multi-indices of the length between $0$ and $n$. Let us denote $e_A=(e^A)^{-1}$.

Let us denote the elements
\begin{eqnarray*}
e,\quad h^{ab}:=h^a h^b, \quad 1\leq a<b\leq n;\quad \ldots,\quad h^{1\ldots n}:=h^1 \cdots h^n
\end{eqnarray*}
by $h^A$, where $A$ is an arbitrary ordered multi-index of the length between $0$ and $n$. Let us consider the following sums over all such $2^n$ multi-indices $A$
\begin{eqnarray}
\sum_A h^A(x) F e_A\label{svert}
\end{eqnarray}
where $F$ is an arbitrary element of the basis (\ref{eebasis}).

Let us consider an arbitrary point $x_0\in\Omega$. By generalized Pauli's theorem \cite{Shirokov}, we have at least one basis element from
(\ref{eebasis}) (we denote it by $F_h$) such that
\begin{equation}
T_h:=\sum_A h^A(x_0) F_h e_A\neq 0. \label{Texpr}
\end{equation}
We define the norm of Clifford algebra elements by
$$
|U|=\sqrt{\Tr(U^\dagger U)},
$$
where $\Tr:\cl^\F(p,q)\to\cl^\F_0(p,q)$ is the projection operation onto the subspace $\cl^\F_0(p,q)$ and the operation of Hermitian conjugation $\dagger$ is defined in \cite{AACA:March:Shir}. Using (\ref{Texpr}), we get
$$
|T_h|=|\sum_A h^A(x_0) F_h e_A|=\delta_h>0.
$$
Since a linear combination of functions of the class $C^k(\Omega)$ is a function of the class $C^k(\Omega)$, we conclude that $|\sum_A h^A(x) F_h e_A|$, $x\in\Omega$, is a continuous function. Thus there exists a real number $\varepsilon_h>0$ such that
$$
|\sum_A h^A(x) F_h e_A|>\delta_h/2,\quad\forall x\in
O_{\varepsilon_h}(x_0).
$$
Consequently, we construct a function
$$
T_h(x)=\sum_A h^A(x) F_h e_A\neq0,\quad \forall x\in
O_{\varepsilon_h}(x_0)
$$
of the class $C^k(O_{\varepsilon_h}(x_0))$. By generalized Pauli's theorem \cite{Shirokov} we have
$$
e^a=T_h^{-1}(x) h^a(x)T_h(x),\quad a=1,\ldots,n,\quad\forall x\in
O_{\varepsilon_h}(x_0)).
$$
Acting the same way (replacing the symbol $h$ by $g$), we can obtain the connection between the elements $e^a$ and $g^a$ using the element $T_g(x)$:
$$
e^a=T_g^{-1}(x) g^a(x)T_g(x),\quad a=1,\ldots,n,\quad\forall x\in
O_{\varepsilon_g}(x_0)).
$$
Choosing $\varepsilon=\min(\varepsilon_h, \varepsilon_g)$, we get
$$
g^a(x)=T^{-1}(x) h^a(x)T(x),\quad a=1,\ldots,n,\quad\forall x\in
O_{\varepsilon}(x_0)),
$$
where $T(x)=T_h(x) T_g^{-1}(x)$. The theorem is proved. $\blacktriangleleft$
\end{proof}

Note that a key role in the proof of the local Pauli's theorem is played by an algorithm \cite{Shirokov} for computing the element, which connects two sets of anticommuting elements by similarity transformation.

Let us formulate and prove the corresponding theorem for the case of odd $n$.

\begin{lemma}\label{lemma1} Let $n$ be a positive odd number and $h^a=h^a(x)$, $g^a=g^a(x)$, $a=1,\dots,n$ be functions $\Omega\to\cl^\R(p,q)$ of the class $C^k(\Omega)$ such that
\begin{eqnarray}
&&h^a(x) h^b(x) + h^b(x) h^a(x)=2\eta^{ab}e, \quad a,b=1,\ldots,n,\quad \forall
x\in\Omega,\nonumber\\
&&g^a(x) g^b(x) + g^b(x) g^a(x)=2\eta^{ab}e, \quad a,b=1,\ldots,n,\quad \forall
x\in\Omega.\nonumber
\end{eqnarray}
Then the products
$$h^{1\ldots n}:=h^{1}(x) h^2(x)\ldots h^n(x),\qquad g^{1\ldots n}:=g^1(x) g^2(x) \ldots g^n(x)$$
do not depend on $x$ and equal $\pm e^{1\ldots n}$ or $\pm e$ (last case is possible only for $p-q=1\mod 4$).
\end{lemma}
\begin{proof}  Lemma \ref{lemma1} follows from the algebraic Pauli's theorem (see \cite{Shirokov}). $\blacktriangleleft$
\end{proof}

\begin{theorem}{\rm (The case of odd $n$ and real Clifford algebra).}\label{th2}

Under the assumptions of Lemma \ref{lemma1}, for any point $x_0\in\Omega$ there exist $\varepsilon>0$ and $T=T(x) : O_\varepsilon(x_0)\to\cl^\R(p,q)$ such that
\begin{enumerate}
\item $T(x)$ is a function of the class $C^k(O_\varepsilon(x_0))$;
\item $T(x)$ is an invertible element of $\cl^\R(p,q)$ for any point $x\in O_\varepsilon(x_0)$;
\item \begin{enumerate}
        \item $g^a(x)=T^{-1}(x)h^a(x)T(x)\quad \Leftrightarrow \quad h^{1\ldots n}=g^{1\ldots n},$
        \item $g^a(x)=-T^{-1}(x)h^a(x)T(x)\quad \Leftrightarrow \quad h^{1\ldots n}=-g^{1\ldots n},$
        \item $g^a(x)=e^{1\ldots n}T^{-1}(x)h^a(x)T(x)\quad  \Leftrightarrow \quad h^{1\ldots n}=e^{1\ldots n} g^{1\ldots n},$
        \item $g^a(x)=-e^{1\ldots n}T^{-1}(x)h^a(x)T(x)\quad  \Leftrightarrow \quad h^{1\ldots n}=-e^{1\ldots n}g^{1\ldots n},$
      \end{enumerate}
where equalities hold for $a=1,\ldots,n$ and $\forall x\in O_\varepsilon(x_0)$;

\item The function $T(x)$ is defined up to multiplication by elements $\lambda(x) e + \nu(x) e^{1\ldots n}$, where $\lambda(x)$ and $\nu(x)$ are real functions of the class $C^k(O_\varepsilon(x_0))$ such that $\lambda(x) e + \nu(x) e^{1\ldots n}$ is an invertible element for any point of the domain $O_\varepsilon(x_0)$.
\end{enumerate}
\end{theorem}

\begin{lemma}\label{lemma2} Let $n$ be a positive odd number and $h^a=h^a(x)$, $g^a=g^a(x)$, $a=1,\dots,n$ be functions $\Omega\to\cl^\C(p,q)$ of the class $C^k(\Omega)$ such that
\begin{eqnarray}
&&h^a(x) h^b(x) + h^b(x) h^a(x)=2\eta^{ab}e, \quad a,b=1,\ldots,n,\quad \forall
x\in\Omega,\nonumber\\
&&g^a(x) g^b(x) + g^b(x) g^a(x)=2\eta^{ab}e, \quad a,b=1,\ldots,n,\quad \forall
x\in\Omega.\nonumber
\end{eqnarray}
Then the products
$$h^{1\ldots n}:=h^{1}(x) h^2(x)\ldots h^n(x),\qquad g^{1\ldots n}:=g^1(x) g^2(x) \ldots g^n(x)$$
do not depend on $x$ and equal $\pm e$ (in the case $p-q=1\mod 4$), $\pm ie$ (in the case $p-q=3\mod 4$), or $\pm e^{1\ldots n}$ (in both cases).
\end{lemma}

\begin{proof} Lemma \ref{lemma2} follows from the algebraic Pauli's theorem (see \cite{Shirokov}). $\blacktriangleleft$
\end{proof}

\begin{theorem}{\rm (The case of odd $n$ and complexified Clifford algebra).}\label{th3}

Under the assumptions of Lemma \ref{lemma2}, for any point $x_0\in\Omega$ there exist $\varepsilon>0$ and $T=T(x) : O_\varepsilon(x_0)\to\cl^\C(p,q)$ such that
\begin{enumerate}
\item $T(x)$ is a function of the class $C^k(O_\varepsilon(x_0))$;
\item $T(x)$ is an invertible element of $\cl^\C(p,q)$ for any point $x\in O_\varepsilon(x_0)$;
\item \begin{enumerate}
        \item $g^a(x)=T^{-1}(x)h^a(x)T(x)\quad \Leftrightarrow \quad h^{1\ldots n}=g^{1\ldots n},$
        \item $g^a(x)=-T^{-1}(x)h^a(x)T(x)\quad \Leftrightarrow \quad h^{1\ldots n}=-g^{1\ldots n},$
        \item $g^a(x)=e^{1\ldots n}T^{-1}(x)h^a(x)T(x)\quad  \Leftrightarrow \quad h^{1\ldots n}=e^{1\ldots n}g^{1\ldots n},$
        \item $g^a(x)=-e^{1\ldots n}T^{-1}(x)h^a(x)T(x)\quad  \Leftrightarrow \quad h^{1\ldots n}=-e^{1\ldots n}g^{1\ldots n},$
        \item $g^a(x)=ie^{1\ldots n}T^{-1}(x)h^a(x)T(x)\quad  \Leftrightarrow \quad h^{1\ldots n}=ie^{1\ldots n}g^{1\ldots n},$
        \item $g^a(x)=-ie^{1\ldots n}T^{-1}(x)h^a(x)T(x)\quad  \Leftrightarrow \quad h^{1\ldots n}=-ie^{1\ldots n}g^{1\ldots n},$
      \end{enumerate}
where equalities hold for $a=1,\ldots,n$ and $\forall x\in O_\varepsilon(x_0)$;

\item The function $T(x)$ is defined up to multiplication by $\lambda(x) e + \nu(x) e^{1\ldots n}$, where $\lambda(x)$ and $\nu(x)$ are complex functions of the class $C^k(O_\varepsilon(x_0))$ such that $\lambda(x) e + \nu(x) e^{1\ldots n}$ is an invertible element for any point of the domain $O_\varepsilon(x_0)$.
\end{enumerate}
\end{theorem}

\begin{proof} Proofs of Theorems \ref{th2} and \ref{th3} are similar to the proof of Theorem \ref{th1} and we must use generalized Pauli's theorems \cite{Shirokov} for the Clifford algebra with odd $n$.

First we prove the connection between the set $h^a=h^a(x)$ and the set $e^a$, $a=1, \ldots, n$, which does not depend on $x$. Instead of (\ref{svert}) we consider the following expressions
$$\sum_{A: |A|=0\mod 2}h^A(x) F_h e_A$$
where we have a sum over ordered multi-indices of even length $|A|$.

The element $F_h$ does not depend on $x$ because it is always among the basis elements $\{e^B\}$ or among the expressions $\{e^B+ e^C\}$ (see \cite{Shirokov}). All other considerations are similar to the considerations for the case of even $n$.

Using the connection between the sets $h^a(x)$ and $e^a$, the sets $g^a(x)$ and $e^a$, we obtain the connection between the sets $g^a(x)$ and $h^a(x)$ in some neighborhood of the point $x_0\in\Omega$. The theorem is proved. $\blacktriangleleft$
\end{proof}

Note that the connection between the sets in the case of odd $n$ (see Theorems \ref{th2} and \ref{th3}) can be written for all cases in the following form:
$$g^{a}(x)=h^{1\ldots n} g_{1\ldots n}T^{-1}(x) h^a(x) T(x),\qquad a=1, \ldots, n,\qquad \forall x\in O_\varepsilon(x_0),$$
where $g_{1\ldots n}:=(g^{1\ldots n})^{-1}$.

\section{Local generalized Pauli's theorem in the entire Euclidean space and the connection with one field equation}

As was shown above, the connection between two sets of elements satisfying the defining anticommutative relations of Clifford algebra is realized in the form of similarity transformation (or in other similar forms in the case of odd $n$) locally, in the neighborhood of the corresponding point of Euclidean space. Does \emph{local generalized Pauli's theorem} hold in the entire Euclidean space? Namely, does there exists a function $T=T(x)$ from the statements of Theorems \ref{th1}, \ref{th2}, and \ref{th3}, which is invertible, continuous, and connects two sets of elements $h^a(x)$, $g^a(x)$, $a=1,\ldots, n$ for any point $x\in V$? In this section we prove the corresponding theorem under additional assumptions (see conditions (\ref{adcond})).

The Cartesian coordinates of Euclidean space $V$, $\dim V=r$ are denoted by $x^\mu$, $\mu=1, \ldots, r$ and the partial derivatives are denoted by $\partial_\mu=\frac{\partial}{\partial x^\mu}$, $\mu=1, \ldots, r$. All the given functions in this section are smooth to simplify the presentation.

Let a set of smooth functions $h^a: V \to\cl^\F(p,q)$, $a=1, \ldots, n$, satisfies
\begin{eqnarray}
h^a(x) h^b(x)+h^b(x) h^a(x)=2\eta^{ab}e,\qquad a, b=1, \ldots, n,\qquad \forall x\in V.\label{hnab}
\end{eqnarray}
In the case of odd $n$, we also require the additional condition
\begin{eqnarray}
\Tr(h^{1\ldots n})=0,\label{adcond}
\end{eqnarray}
where $\Tr:\cl^\F(p,q)\to\cl^\F_0(p,q)$ is the projection operation onto the subspace $\cl^\F_0(p,q)$. We need this condition (\ref{adcond}) to obtain independent elements $h^a(x)$, $a=1, \ldots, n$, which generate the basis of the Clifford algebra $\cl^\F(p,q)$. Otherwise, they can generate a basis of Clifford algebra of lower dimension (see \cite{Shirokov}). Under this additional condition, we have only two cases of connection between two sets of Clifford algebra elements in the case of odd $n$: $g^a=\pm T^{-1} h^a T$ (instead of four and six cases as in Theorems \ref{th2} and \ref{th3}).

Note that the problem of connection between the sets $h^a(x)$ and $g^a(x)$, $a=1, \ldots, n$ is equivalent to the problem of connection between the set $h^a(x)$, $a=1, \ldots, n$ and the set of generators $e^a$, $a=1, \ldots, n$, that do not depend on $x\in V$. Using the connection between the sets $h^a(x)$ and $e^a$, the sets $g^a(x)$ and $e^a$, we obtain the connection between the sets $h^a(x)$ and $g^a(x)$. Therefore, in what follows we  consider the problem of connection between the set $h^a(x)$, $a=1, \ldots, n$ with conditions (\ref{hnab}), (\ref{adcond}) and the set $e^a$, $a=1, \ldots, n$.

In the papers \cite{marchuk1}, \cite{ofe}, \cite{ccYM}, one primitive field equation (the system of partial differential equations) for the spin connection of the general form was considered. In \cite{marchuk1} and \cite{ofe}, the expressions $h^\mu$ were considered as vector expressions with respect to the orthogonal transformations of coordinates. In the present paper, as well as in \cite{ccYM}, we consider instead of them expressions $h^a$, which do not change under orthogonal transformations of coordinates.

Consider the following field equation (system of equations)
\begin{eqnarray}
\partial_\mu h^a-[C_\mu, h^a]=0,\qquad a=1, \ldots, n,\qquad \mu=1, \ldots, r,\label{prfe2}
\end{eqnarray}
where the components $C_\mu: V\to\cl^\F(p,q)$ of covector field with values in the Clifford algebra are considered as unknowns. Thus, we have $n\times r$ equations for $r$ unknown functions.

Since the expressions $C_\mu$ are inside the commutator in (\ref{prfe2}), then it is convenient to consider these expressions up to element of the Clifford algebra center: $C_\mu: V\to \cl^\F(p,q)\setminus \Cen(\cl^\F(p,q))$, where $$
\Cen(\cl^\F(p,q))=\left\{\begin{array}{ll}
\cl^\F_0(p,q), & \hbox{in the case of even $n$;} \\
\cl^\F_0(p,q)\oplus\cl^\F_n(p,q), & \hbox{in the case of odd $n$.}
\end{array}
\right.
$$

The equation (\ref{prfe2}) is gauge invariant. Namely, the following expressions
\begin{eqnarray}
\acute h^a =S^{-1}h^a S,\qquad \acute C_\mu=S^{-1}C_\mu S-S^{-1}\partial_\mu S\label{preobr}
\end{eqnarray}
for a smooth invertible function $S: V\to \cl^\F(p,q)$ such that $S^{-1}\partial_\mu S:V\to \cl^\F(p,q)\setminus \Cen(\cl^\F(p,q))$,
satisfy the system of equations of the same form
$$\partial_\mu\acute h^a-[\acute C_\mu , \acute h^a]=0,\qquad a=1, \ldots, n,\qquad \mu=1, \ldots, r.$$

In \cite{ofe}, \cite{ccYM}, it is proved that the system (\ref{prfe2}) has a unique solution $C_\mu: V\to \cl^\F(p,q)\setminus \Cen(\cl^\F(p,q))$ of the following form:
\begin{equation}
C_\mu=\sum_{k=1}^{2[\frac{n}{2}]} \mu_k \pi[h]_k ((\partial_\mu h^a) h_a),\qquad \mu_k=\frac{1}{n-(-1)^k(n-2k)},\qquad h_a:=(h^a)^{-1},\label{resh}
\end{equation}
where $\pi[h]_k: \cl^\F(p,q) \to \cl^\F[h]_k(p,q)$ is the projection operation onto the subspace $\cl^\F[h]_k(p,q)$ spanned over the basis elements $h^{a_1 \ldots a_k}$ with ordered multi-indices of length $k$. Explicit expressions for the elements $C_\mu$ in the cases of small $n$ are given in \cite{ofe} and \cite{ccYM}. We say that the solution (\ref{resh}) describes \emph{the spin connection of general form}.

If the generators $e^a$ of Clifford algebra, which do not depend on $x$, are considered as the expressions $\acute h^a$ (\ref{preobr}), then the corresponding connection for these elements given by the equations (\ref{prfe2}) is equal to zero: $\acute C_\mu=0$. Using (\ref{preobr}), we get
\begin{eqnarray}
\partial_\mu S(x)=C_\mu(x) S(x),\qquad \mu=1, \ldots, r.\label{syst0}
\end{eqnarray}
The system of partial differential equations (\ref{syst0}) with known $C_\mu(x)$ is considered as a system for finding a function $S=S(x)$, which is invertible in the entire Euclidean space and connects two sets $h^a(x)$ and $e^a$, $a=1, \ldots, n$. We have the following theorem.

\begin{theorem}{\rm (Local Pauli's theorem in the entire Euclidean space).}\label{th4}

Let us consider functions $h^a: V \to\cl^\F(p,q)$, $a=1, \ldots, n$, that satisfy
\begin{eqnarray}
h^a(x) h^b(x)+h^b(x) h^a(x)=2\eta^{ab}e,\qquad a, b=1, \ldots, n,\qquad \forall x\in V\nonumber
\end{eqnarray}
and, in the case of odd $n$, additional condition $\Tr(h^{1\ldots n})=0$.

Then there exists a function $S=S(x): V\to\cl^\F(p,q)$, $\exists \, S^{-1}(x) \, \forall x\in V$, satisfying the following system of equations
\begin{eqnarray}
\partial_\mu S(x)=C_\mu(x) S(x),\qquad \mu=1, \ldots, r\label{syst}
\end{eqnarray}
for all $x \in V$, where $C_\mu: V\to \cl^\F(p,q)\setminus \Cen(\cl^\F(p,q))$ is a unique solution of the system of equations
\begin{eqnarray}
\partial_\mu h^a-[C_\mu, h^a]=0,\qquad a=1, \ldots, n,\qquad \mu=1, \ldots, r,\label{prfe}
\end{eqnarray}
and there exists a function $T(x)=S(x) K$ (for some invertible element $K\in\cl^\F(p,q)$, which does not depend on $x$), which is also invertible in the entire Euclidean space solution of the system (\ref{syst}) and connects two sets of elements:
\begin{eqnarray}
e^a=T^{-1}(x)h^a(x) T(x),\qquad a=1, \ldots, n,\qquad\forall x\in V\label{b1}
\end{eqnarray}
in the case of even $n$ and
\begin{eqnarray}
e^a=h^{1\ldots n}e_{1\ldots n} T^{-1}(x)h^a(x) T(x),\qquad a=1, \ldots, n,\qquad \forall x\in V\label{b2}
\end{eqnarray}
in the case of odd $n$, where $h^{1\ldots n}e_{1\ldots n}=\pm e$.
\end{theorem}

\begin{proof} Using methods of differential geometry, it can be proved that the function $S(x)$ from the statement of the theorem always exists.

In the papers \cite{ofe}, \cite{ccYM}, it is proved that from the system of equations
\begin{eqnarray}
\partial_\mu h^a-[C_\mu, h^a]=0,\qquad a=1, \ldots, n,\qquad \mu=1, \ldots, r\label{tt}
\end{eqnarray}
the following relation for the functions $C_\mu$ follows
\begin{eqnarray}
\partial_\mu C_\nu-\partial_\nu C_\mu-[C_\mu, C_\nu]=0,\qquad \mu, \nu=1, \ldots, r,\qquad \forall x\in V.\label{collor}
\end{eqnarray}
In terms of differential geometry this means that the curvature
$$R_{\mu\nu}:=\partial_\mu C_\nu-\partial_\nu C_\mu-[C_\mu, C_\nu]$$
is equal to zero, i.e. the connection $C_\mu$ is flat. It is known that every flat connection on a simply-connected manifold is trivial, i.e. it can be represented in the form $C_\mu=(\partial_\mu S) S^{-1}$ for some function $S=S(x)$ (see, for example, \cite{conn}). Since Euclidean space is simply-connected, there exists invertible in the entire Euclidean space function $S(x)$, which satisfies (\ref{syst}). Substituting the expression $C_\mu=(\partial_\mu S)S^{-1}$ into the equation (\ref{prfe}), we get
\begin{eqnarray}
\partial_\mu h^a-(\partial_\mu S)S^{-1}h^a+h^a (\partial_\mu S)S^{-1}=0.\label{g5}
\end{eqnarray}
Using $S S^{-1}=e$, we obtain
\begin{eqnarray}
(\partial_\mu S)S^{-1}+S \partial_\mu(S^{-1})=0.\label{g6}
\end{eqnarray}
Multiplying both sides of the equation (\ref{g5}) from the left by $S^{-1}$, from the right by $S$, and using (\ref{g6}), we get
\begin{eqnarray}
\partial_\mu(S^{-1}h^a S)=0,\qquad \mu=1, \ldots, r.\label{partial}
\end{eqnarray}
Since (\ref{partial}), it follows that the set
$$f^a:=S^{-1}h^a S,\qquad a=1, \ldots, n$$
does not depend on $x\in V$ and satisfy the defining anticommutative conditions of Clifford algebra:
$$f^a f^b+f^b f^a=S^{-1}h^a S S^{-1}h^b S+S^{-1}h^b SS^{-1}h^a S=h^a h^b+h^b h^a=2\eta^{ab}e.$$
If the function $S(x)$ is a solution of the system (\ref{syst}), then any function of the form $S(x)K$, for the independent on $x$ element $K$, is also a solution of the system (\ref{syst}). By the algebraic generalized Pauli's theorem \cite{Shirokov}, there is an invertible element $K\in\cl^\F(p,q)$ such that
$$e^a=K^{-1}f^a K,\qquad a=1, \ldots, n$$
in the case of even $n$ and
$$e^a=f^{1\ldots n}e_{1\ldots n}K^{-1}f^a K,\qquad a=1, \ldots, n$$
in the case of odd $n$. We conclude that the element
$$T(x)=S(x)K$$
connects two sets of elements $e^a$ and $h^a(x)$, $a=1, \ldots, n$ in the form (\ref{b1}) and (\ref{b2}). The theorem is proved. $\blacktriangleleft$
\end{proof}

Note that Theorem \ref{th4} gives us an algorithm for computing the function $S=S(x)$. Using this algorithm and algorithm for computing the element $K$ from the algebraic Pauli's theorem (see \cite{Shirokov}), we obtain an algorithm for computing the function $T(x)=S(x)K$, which connects two sets of elements $h^a(x)$, $e^a$, $a=1, \ldots, n$.

Below we also give two particular cases (Theorems \ref{th5} and \ref{th6}) of the statement of Theorem \ref{th4}, in which the function $T(x)$ has a simpler form. We give another proofs of these theorems without using the fact that any flat connection on a simply-connected manifold is trivial. In this particular cases, it is sufficient to use the theory of matrix differential equations or the Poincare lemma.

The following theorem describes the local Pauli's theorem in the case of Euclidean space $V=\R^1$ of dimension $r=\dim V=1$.

\begin{theorem} {\rm (Particular case: $r=1$, $n\geq 2$).} \label{th5}

Let us consider smooth functions $h^a: \R \to\cl^\F(p,q)$, $a=1, \ldots, n$, that satisfy
\begin{eqnarray}
h^a(x) h^b(x)+h^b(x) h^a(x)=2\eta^{ab}e,\qquad a, b=1, \ldots, n,\qquad \forall x\in \R \nonumber
\end{eqnarray}
and additional condition $\Tr(h^{1\ldots n})=0$ in the case of odd $n$.

Then there exists a function $T=T(x): \R \to \cl^\F(p,q)$ such that
\begin{eqnarray}
e^a=T^{-1}(x)h^a(x) T(x),\qquad a=1, \ldots, n,\qquad\forall x\in \R\label{svyaz7}
\end{eqnarray}
in the case of even $n$ and
\begin{eqnarray}
e^a=h^{1\ldots n}e_{1\ldots n} T^{-1}(x)h^a(x) T(x),\qquad a=1, \ldots, n,\qquad \forall x\in \R\label{svyaz27}
\end{eqnarray}
in the case of odd $n$, where $h^{1\ldots n}e_{1\ldots n}=\pm e$.

Moreover, $T(x)=S(x) K$, where $S(x)$ is any invertible in the entire Euclidean space solution of the equation
\begin{eqnarray}
\frac{dS(x)}{dx}=C_1(x) S(x),\label{syst7}
\end{eqnarray}
$C_1: \R\to \cl^\F(p,q)\setminus \Cen(\cl^\F(p,q))$ is a unique solution of the system of differential equations
\begin{eqnarray}
\frac{dh^a}{dx}-[C_1, h^a]=0,\qquad a=1, \ldots, n,\label{prfe7}
\end{eqnarray}
and $K$ is an invertible element of the Clifford algebra $\cl^\F(p,q)$.
\end{theorem}
\begin{proof}
In the case $r=\dim V=1$, the system of partial differential equations (\ref{syst}) becomes the ordinary differential equation (\ref{syst7}) and the system of partial differential equations (\ref{prfe}) becomes the system of ordinary differential equations (\ref{prfe7}).

According to the theory of matrix differential equations (see, for example, \cite{Gant}, Section 14), the equation (\ref{syst7}) for a continuous function $C_1(x)$ has always invertible in the entire Euclidean space solution $S(x)$, and the general solution of the equation (\ref{syst7}) has the form
$$T(x)= S(x)K,$$
where $K\in\cl^\F(p,q)$ is any element that does not depend on $x$. The function $T(x)=S(x)K$ for some invertible element $K\in\cl^\F(p,q)$ connects two sets of elements in the form (\ref{svyaz7}) in the case of even $n$ and in the form (\ref{svyaz27}) in the case of odd $n$.

Note that solution of the system (\ref{syst7}) in the general case can be written in the form of multiplicative integral (see \cite{Gant}). In the case of additional conditions $[C_1(x_1),C_1(x_2)]=0$ for any $x_1, x_2\in\R$, the solution has a simpler form
$$S(x)=\exp(\int_{x_0}^x C_1(x) dx)$$
for some point $x_0\in\R$. $\blacktriangleleft$
\end{proof}

Now let us consider for arbitrary $r\geq 1$ the case when the functions (\ref{hnab}) take values in the subspace $\cl_1^\F(p,q)$ spanned over the generators $e^a$, $a=1, \ldots, n$:
\begin{eqnarray}
h^a(x): V\to \cl^\F_1(p,q)\label{dop2}
\end{eqnarray}
In this case we have
$$h^a(x)=y^a_b(x) e^b$$
for some smooth functions $y^a_b(x): V\to\F$. The conditions
$$h^a(x) h^b(x)+h^b(x) h^a(x)=2\eta^{ab}e,\qquad a, b=1, \ldots, n,\qquad \forall x\in V$$
on the functions $h^a(x)$, $a=1, \ldots, n$ are equivalent to the following conditions on the functions $y^a_b(x)$, $a, b=1, \ldots, n$:
\begin{eqnarray}
y^a_b(x) y^c_d(x) \eta^{bd}=\eta^{ac},\qquad \forall x\in V.\label{tetr}
\end{eqnarray}
Note that (\ref{tetr}) is the orthogonality condition for the matrix
$$Y=||y^a_b||\in\Or(p,q, \F)=\{Y\in\Mat(n, \F), Y^\T \eta Y=\eta\},$$
where $Y=||y^a_b||$ means that the entry in the $a$-th row and $b$-th column of the matrix $Y$ is denoted by $y^a_b$.

In the case $h^a: V\to \cl^\F_1(p,q)$, the unique solution (\ref{resh}) of the system of equations
$$\partial_\mu h^a-[C_\mu, h^a]=0,\qquad a=1, \ldots, n,\qquad \mu=1, \ldots, r$$
has the form (see \cite{marchuk1})
\begin{eqnarray}
C_\mu=\frac{1}{4}(\partial_\mu h^a) h_a,\qquad h_a:=(h^a)^{-1},\label{spincon}
\end{eqnarray}
which is known as \emph{the spin connection} (see, for example, \cite{GShW}). In \cite{marchuk1}, it is shown that the function (\ref{spincon}) takes values in the subspace of grade 2 in this case, i.e. $C_\mu: V\to \cl^\F_2(p,q)$.

Note that the subspace $\cl^\R_2(p,q)$ is the Lie algebra (with respect to the commutator) of the spin group $\Spin_+(p,q)$ and the exponents of the elements of $\cl^\R_2(p,q)$ are elements of the group $\Spin_+(p,q)$ \cite{Lounesto}, \cite{MarchukShirokovbook}. The formula $S^{-1} e^a S=y^a_b e^b=:h^a$ describes two-sheeted covering of the orthogonal group $\SO_+(p,q)$ by the corresponding spin group $\Spin_+(p,q)$: for each orthogonal matrix $Y=||y^a_b||\in\SO_+(p,q)$ there exist two elements $\pm S\in\Spin_+(p,q)$ of the corresponding spin group.

\medskip

The following theorem describes the local Pauli's theorem in the entire Euclidean space in the case $n=2$ and an arbitrary $r$ with the additional assumption $h^a: V\to \cl^\F_1(p,q)$, $a=1, 2.$

\begin{theorem} {\rm (Particular case: $n=2$, $r\geq 1$).}\label{th6}

Let us consider smooth functions $h^a: V\to \cl^\F_1(p,q)$, $a=1, 2$ with values in the subspace of grade $1$ that satisfy
\begin{eqnarray}
h^a(x) h^b(x)+h^b(x) h^a(x)=2\eta^{ab}e,\qquad a,b=1,2,\qquad \forall x\in V.\label{soot}
\end{eqnarray}
Then there exists a function $C(x): V\to \cl_2^\F(p,q)$ such that
$$dC(x)=C_1(x)dx^1+\cdots+C_r(x)dx^r,\qquad C_\mu(x)=\frac{1}{4}(\partial_\mu h^a) h_a.$$
Moreover, the function
$$T(x)=\exp(C(x))K$$
satisfies
$$e^a=T^{-1}(x) h^a(x) T(x),\qquad a=1, 2,\qquad \forall x\in V,$$
for some invertible element of Clifford algebra $K\in\cl^\F(p,q)$.
\end{theorem}

\begin{proof} As mentioned before the theorem, if $h^a: V\to \cl^\F_1(p,q)$, then the functions $C_\mu$, $\mu=1, \ldots, r$ take values in the subspace $\cl^\F_2(p,q)$ (see \cite{marchuk1}).

In the case $n=2$, all elements of the subspace $\cl^\F_2(p,q)$ have the form $\lambda e^{12}$, $\lambda \in \F$. Hence all the functions $C_\mu:V\to\cl^\F_2(p,q)$ commute with each other
\begin{eqnarray}
[C_\mu(x), C_\nu(x)]=0,\qquad \mu, \nu=1, \ldots, r,\qquad \forall x\in V.\label{commut}
\end{eqnarray}
As mentioned above, in the papers \cite{ofe}, \cite{ccYM}, it is proved that from the system of equations (\ref{tt}) follows
(\ref{collor}). Thus, by (\ref{collor}), the conditions (\ref{commut}) are equivalent to
\begin{eqnarray}
\partial_\mu C_\nu=\partial_\nu C_\mu,\qquad \mu, \nu=1, \ldots, r,\qquad \forall x\in V.\label{prt}
\end{eqnarray}

Let us consider the following 1-form $L(x)=C_\mu(x) dx^\mu$. By the Poincare lemma, if this form is closed $dL(x)=0$, then it is exact, i.e. there exists a $0$-form $C(x)$ such that $dC(x)=L(x)$. Since (\ref{prt}), it follows that there exists a function $C(x): V\to\cl^\F_2(p,q)$ such that
$$dC(x)=C_1(x) dx^1+\cdots+C_r(x) dx^r.$$
Thus, under the conditions (\ref{prt}), the system of equations (\ref{syst}) can be written in the form
\begin{eqnarray}
\partial_\mu S(x)=\partial_\mu (C(x)) S(x),\qquad \mu=1, \ldots, r.\label{syst9}
\end{eqnarray}
Let us consider the expression
\begin{eqnarray}
\exp(C(x))=e+C(x)+\frac{1}{2!}C^2(x)+\frac{1}{3!}C^3(x)+\cdots=\sum_{k=0}^\infty \frac{1}{k!}C^k(x).\nonumber
\end{eqnarray}
This series is always convergent and invertible (see, for example \cite{Gant})
\begin{eqnarray}
(\exp(C(x)))^{-1}=\exp(-C(x)).\label{invexp}
\end{eqnarray}
Since $C_\mu(x):V\to\cl^\F_2(p,q)$, we conclude that $C(x)$ also takes values in $\cl_2^\F(p,q)$. The functions $C^k(x)$, $k=1, 2, \ldots$, and $\exp(C(x))$ take values in $\cl_0^\F(p,q)\oplus\cl_2^\F(p,q)$. In the Clifford algebra of dimension $n=2$ we have
$[\cl^\F_{0}(p,q)\oplus\cl^\F_{2}(p,q), \cl^\F_{2}(p,q)]=0$, and hence
$$[C^k(x), \partial_\mu C(x)]=0,\qquad \forall k=1, 2, \ldots,$$
which is a sufficient condition for the following formula (see \cite{Gant})
\begin{eqnarray}
\partial_\mu(\exp(C(x)))=(\partial_\mu C(x))\exp(C(x)).\label{proiz}
\end{eqnarray}
From (\ref{invexp}) and (\ref{proiz}), it follows that the function
\begin{eqnarray}
S(x)=\exp(C(x))K\nonumber
\end{eqnarray}
is an invertible in the entire Euclidean space solution of the system (\ref{syst9}) for any invertible element $K\in\cl^\F(p,q)$ that does not depend on $x$:
$$\partial_\mu (\exp(C(x))K)=\partial_\mu (\exp(C(x)))K=(\partial_\mu C(x))\exp(C(x))K.$$
Using the algebraic Pauli's theorem, we conclude that for some invertible $K\in\cl^\F(p,q)$ the function $T(x)=\exp(C(x))K$  connects two sets:
$$e^a=T^{-1}(x) h^a(x) T(x),\qquad a=1, 2,\qquad \forall x\in V.$$
The theorem is proved. $\blacktriangleleft$
\end{proof}

%

Let us give some examples illustrating the statement of Theorem \ref{th6}.

1) Let us consider the real Clifford algebra $\cl^\R(2,0)$ and the functions $h^a: V\to \cl^\F_1(2,0)$ satisfying the relations (\ref{soot}), $h^a=y^a_b e^b$. We can parameterize the elements of the matrix $Y=||y^a_b||\in\Or(2)$ by the function $\varphi=\varphi(x):V\to\R$. We have two cases ($\det Y=\pm 1$):
\begin{eqnarray}
&&h^1=\cos\varphi \,e^1+\sin\varphi \,e^2,\qquad h^2=-\sin\varphi \,e^1+\cos\varphi \,e^2;\label{1sl}\\
&&h^1=\cos\varphi \,e^1+\sin\varphi \,e^2,\qquad h^2=\sin\varphi \,e^1-\cos\varphi \,e^2.\label{2sl}
\end{eqnarray}
In both cases, after direct calculations, we obtain
$$C_\mu=\frac{1}{4}(\partial_\mu h^a) h_a=-\frac{\partial_\mu \varphi}{2}e^{12}.$$
We conclude that there exists a function $C(x)=-\frac{\varphi(x)}{2}e^{12}$ such that $C_\mu(x) dx^\mu=dC(x)$.
We obtain the following solution of the equation $\partial_\mu S(x)=C_\mu(x) S(x)$: $$S(x)=\exp(C(x))K=(\cos(\frac{\varphi(x)}{2})\,e-\sin(\frac{\varphi(x)}{2})\,e^{12})K,$$
where $K$ is an arbitrary element of Clifford algebra that does not depend on $x$. We take $T(x)=\exp (C(x))K$ for some invertible element $K$. In the first case (\ref{1sl}), we take $K=e$, in the second case (\ref{2sl}), we take $K=e^{12}$. The element $T(x)$ is invertible for any $x\in V$ and connects two sets of elements $e^a=T^{-1}(x) h^a(x) T(x)$, $\forall x\in V$. Indeed, it is easy to verify
\begin{eqnarray}
&&(\cos \frac{\varphi}{2}e+\sin\frac{\varphi}{2}e^{12})(\cos \varphi \, e^1+\sin \varphi\, e^2)(\cos \frac{\varphi}{2}e-\sin\frac{\varphi}{2}e^{12})=e^1,\nonumber\\
&&(\cos \frac{\varphi}{2}e+\sin\frac{\varphi}{2}e^{12})(-\sin \varphi\, e^1+\cos\varphi\, e^2)(\cos \frac{\varphi}{2}e-\sin\frac{\varphi}{2}e^{12})=e^2.\nonumber
\end{eqnarray}

2) The case of the real Clifford algebra $\cl^\R(0,2)$ is considered similarly. We have two cases: (\ref{1sl}) and (\ref{2sl}). In both cases, we get
$$C_\mu=\frac{1}{4}(\partial_\mu h^a) h_a=\frac{\partial_\mu \varphi}{2}e^{12}$$
and
$$T(x)=\exp(C(x))K=(\cos(\frac{\varphi(x)}{2})\,e+\sin(\frac{\varphi(x)}{2})\,e^{12})K,$$
which satisfies $e^a=T^{-1}(x) h^a(x) T(x)$, $\forall x\in V$ for some invertible $K$.

3) In the case of the real Clifford algebra $\cl^\R(1,1)$ we have four cases (since the group $\Or(1,1)$ has four connected components):
\begin{eqnarray}
&&h^1=\cosh\varphi \,e^1+\sinh\varphi \,e^2,\qquad h^2=\sinh\varphi \,e^1+\cosh\varphi \,e^2;\nonumber\\
&&h^1=\cosh\varphi \,e^1+\sinh\varphi \,e^2,\qquad h^2=-\sinh\varphi \,e^1-\cosh\varphi \,e^2;\nonumber\\
&&h^1=-\cosh\varphi \,e^1-\sinh\varphi \,e^2,\qquad h^2=\sinh\varphi \,e^1+\cosh\varphi \,e^2;\nonumber\\
&&h^1=-\cosh\varphi \,e^1-\sinh\varphi \,e^2,\qquad h^2=-\sinh\varphi \,e^1-\cosh\varphi \,e^2.\nonumber
\end{eqnarray}
In all cases, we have
$$C_\mu=\frac{1}{4}(\partial_\mu h^a) h_a=-\frac{\partial_\mu \varphi}{2}e^{12}$$
and
$$T(x)=\exp(C(x))K=(\cosh(\frac{\varphi(x)}{2})\,e-\sinh(\frac{\varphi(x)}{2})\,e^{12})K,$$
which satisfies $e^a=T^{-1}(x) h^a(x) T(x)$, $\forall x\in V$ for some invertible $K$.

4) Let us consider the case of the real Clifford algebra $\cl^\R(3,0)$ and the functions $h^a: V\to \cl^\F_1(3,0)$ satisfying the relations (\ref{hnab}), $h^a=y^a_b e^b$. For simplicity, we consider the case of the matrix $Y=||y^a_b||\in\Or(3)$ with the determinant $\det Y=1$. In this case, the matrix $Y\in\SO(3)$ can be parameterized by three Euler angles $\varphi(x)$, $\psi(x)$, $\theta(x)$, depending on $x$. We have $0\leq \varphi, \psi < 2\pi$, $0\leq \theta <\pi$.

The matrix $Y$ has the form
$$\left(
  \begin{array}{ccc}
    \cos\varphi\cos\psi\cos\theta-\sin\varphi\sin\psi & -\cos\varphi\sin\psi\cos\theta-\sin\varphi\cos\psi & \cos\varphi\sin\theta \\
    \sin\varphi\cos\psi\cos\theta+\cos\theta\sin\psi & \sin\varphi\sin\psi\cos\theta+\cos\varphi\cos\psi & \sin\varphi\sin\theta \\
    -\cos\psi\sin\theta & \sin\psi\sin\theta & \cos\theta \\
  \end{array}
\right).$$
We get
\begin{eqnarray}
&&C_\mu=\frac{1}{4}(\partial_\mu h^a)h_a= \frac{1}{2}((\cos\theta\,\partial_\mu\varphi+\partial_\mu\psi)e^{12}+\nonumber\\ &&+(-\sin\psi\sin\theta\,\partial_\mu\varphi-\cos\psi\,\partial_\mu\theta )e^{13}+(\cos\psi\sin\theta\,\partial_\mu\varphi+\sin\psi\,\partial_\mu\theta)e^{23}).\nonumber
\end{eqnarray}
In this example, we obtain $[C_\mu, C_\nu]\neq 0$, and therefore $\partial_\mu C_\nu \neq \partial_\nu C_\mu$ and the Poincare lemma is not applicable, unlike the case $n=2$.

\medskip

Problems related to the local Pauli's theorem are useful in field theory, in the study of the Dirac equation \cite{Marchukbook} and the Yang-Mills equations \cite{marchuk1}, \cite{ccYM}. In the paper \cite{ccYM}, a class of covariantly constant solutions of the Yang-Mills equations is proposed. The connection between these solutions and constant solutions is described by the local Pauli's theorem.

An interesting question is whether the Pauli's theorem is valid on the curved manifolds. Our hypothesis is that the Pauli's theorem will be true only in some particular cases (see Theorems \ref{th5} and \ref{th6}) depending on $n$, dimension of the manifold $r$, or topological properties of the manifold (simply connectedness).

\paragraph{{\bf}ACKNOWLEDGEMENTS.} The authors are grateful to N. Hitchin and M. Katanaev for useful discussions. The authors are grateful to the reviewers for their careful reading of the paper
and helpful remarks. This work is supported by the Russian Science Foundation (project 18-71-00010).

\begin{footnotesize}
\begin{flushleft}
Nikolay G. Marchuk\\
\textit{Steklov Mathematical Institute of Russian Academy of Sciences, 119991, Moscow, Russia}\\
\textit{E-mail:} nmarchuk@mi.ras.ru\\
\end{flushleft}
\end{footnotesize}

\begin{footnotesize}
\begin{flushleft}
Dmitry S. Shirokov\\
\textit{National Research University Higher School of Economics, 101000, Moscow, Russia}\\
\textit{Kharkevich Institute for Information Transmission Problems of Russian Academy of Sciences, 127051, Moscow, Russia}\\
\textit{E-mail:} dm.shirokov@gmail.com\\
\end{flushleft}
\end{footnotesize}


\begin{thebibliography}{99}
\bibitem{Gant} F.R. Gantmacher, \textit{The theory of matrices}, GITTL, Moscow, 1953, Russian.
 \bibitem{GShW} M. Green, J. Schwarz, E. Witten, \textit{Superstring theory}, 1, Mir, Moscow, 1990, Russian.
\bibitem{Lounesto} P. Lounesto, \textit{Clifford Algebras and Spinors}, Cambridge Univ. Press, Cambridge, 1997.
\bibitem{AACA:March:Shir} N.G. Marchuk, D.S. Shirokov, \textit{Unitary spaces on Clifford algebras}, Advances in Applied Clifford Algebras, \textbf{18(2)}, 2008.
\bibitem{ofe} N.G. Marchuk, D.S. Shirokov, \textit{General solutions of one class of field equations}, Rep. Math. Phys., \textbf{78(3)}, 2016, 305-326.
\bibitem{MarchukShirokovbook} N.G. Marchuk, D.S. Shirokov, \textit{Introduction to the theory of Clifford algebras}, Phasis, Moscow, 2012, 590 pp., Russian.
\bibitem{marchuk1} N.G. Marchuk, \textit{On a field equation generating a new class of particular solutions to the Yang-Mills equations}, Proceedings of the Steklov Institute of Mathematics, 2014, Vol. \textbf{285}, 197-210.
\bibitem{tensor} N.G. Marchuk, \textit{Demonstration Representation and Tensor Products of Clifford Algebras}, Proc. Steklov Inst. Math., \textbf{290}, 2015, 143-154.
\bibitem{Marchukbook} N. Marchuk, \textit{Field theory equations}, Amazon, CreateSpace, 2012, 290 pp.
\bibitem{Pauli} W. Pauli, \textit{Contributions Mathematiques a la Theorie des Matrices de Dirac}, Ann. Inst. Henri Poincare 6, 1936, 109-136.
\bibitem{ccYM} D.S. Shirokov, \textit{Covariantly constant solutions of the Yang-Mills equations}, Advances in Applied Clifford Algebras, \textbf{28}, 2018, 53.
\bibitem{Shirokov} D.S. Shirokov, \textit{Extension of Pauli's theorem to Clifford algebras}, Dokl. Math., \textbf{84(2)}, 2011, 699-701.
\bibitem{spin} D.S. Shirokov, \textit{Calculations of elements of spin groups using generalized Pauli's theorem}, Advances in Applied Clifford Algebras, \textbf{25(1)}, 2015, 227-244.
\bibitem{tmf} D.S. Shirokov, \textit{Pauli theorem in the description of n-dimensional spinors in the Clifford algebra formalism}, Theoret. and Math. Phys., \textbf{175(1)}, 2013, 454-474.
\bibitem{conn} G. Walschap, \textit{Metric structures in differential geometry}, Series: Graduate Texts in Mathematics, \textbf{224}, Springer, 2004.
\end{thebibliography}
\end{document}